\documentclass[aps,prb,twocolumn,showpacs]{revtex4-1}
\usepackage{graphicx}
\usepackage{bm}
\usepackage{wasysym}

\begin{document}

\title{Spin structure factor and thermodynamics in the antiferromagnetic quantum Ising model in the pyrochlore lattice}
\author{Cheng-Ju Lin$^1$, Chen-Nan Liao$^2$}
\author{Chyh-Hong Chern$^{1}$}
\email{Corresponding author: chchern@ntu.edu.tw}
\affiliation{$^1$Center for Theoretical Sciences and Department of Physics, National Taiwan University, Taipei 10617, Taiwan}
\affiliation{$^2$Department of Industrial Engineering and Operations Research, University of California, Berkeley, CA 94720-1777}

\date{\today}
\begin{abstract}
We numerically compute the temperature dependence of spin structure factor and thermodynamic quantities in the antiferromagnetic quantum Ising model in the pyrochlore lattice.  This model exhibits spin disorder ground state with exponentially-decayed spin correlation.  We reproduce the temperature dependence of the pinch point structure in the neutron scattering experiment and correct entropy obtained from the measurement of the specific heat.
 \end{abstract}
\maketitle
\section{Introduction}\label{section:introduction}

\begin{figure}[htb]
\includegraphics[width=0.4\textwidth]{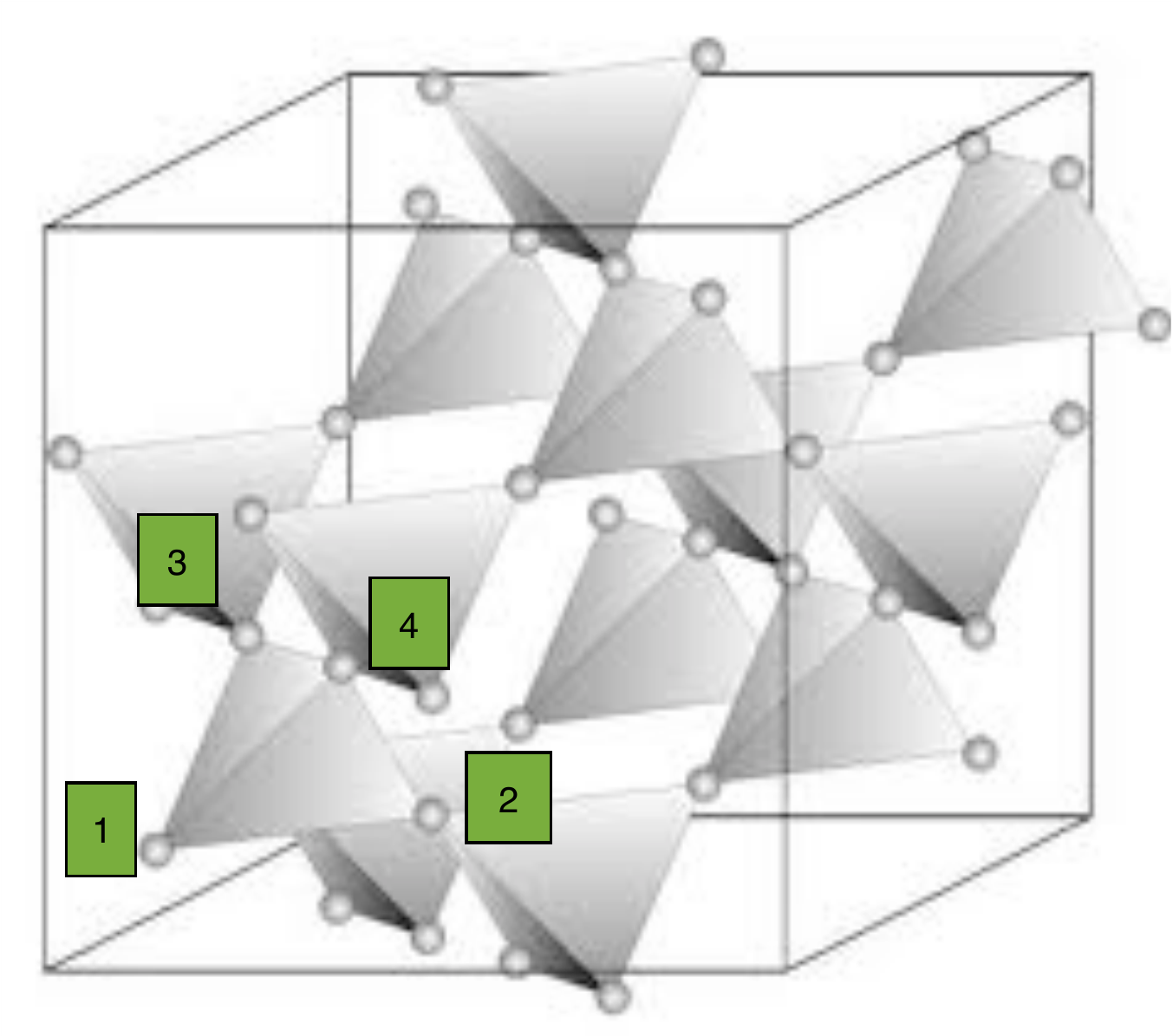}
\caption{(Color online) The cubic unit cell of the pyrochlore lattice.  There are 16 spins containing 4 tetrahedra in a unit cell.}\label{Fig:pyrochlore}
\end{figure}

Searching for magnetic monopoles is a historical and long persisting interest in physics.  This interest was recently invigoratingly aroused by the discovery of the magnetic-monopole-like excitation in the spin ice material.  Magnetic ions in the spin ice material, for example Dy$_2$Ti$_2$O$_7$ and Ho$_2$Ti$_2$O$_7$, form the pyrochlore lattice, a network of corner-shared tetrahedra shown in Fig.~(\ref{Fig:pyrochlore}).  Due to the nature of high spin quantum number\cite{snyder2004}, spin ice material is often regarded as a classical spin system.  The magnetic-monopole-like excitation in spin ice material corresponds to a classical configuration of spin orientation.  Before introducing the monopole, let us review the ground state property first.

The ferromagnetic exchange interaction, crystal field, and the dipolar interaction play the most important roles determining the ground state in the spin ice material.  The anisotropic ferromagnetic exchange and the crystal field restrict the spins to orientate in the local [111] direction, which is along the center of the tetrahedron to the spin site.  The spin configuration of the ground state in one tetrahedron is the "2-in-2-out" structure where two spins point outward the tetrahedron and the other two point inward.  This structure reveals two facts in the spin ice material.  One is the nature of the high spin frustration.  The other is the macroscopic ground state degeneracy.  Since there can be 6 types of ground state configuration in one tetrahedron, the total ground state degeneracy grows in proportion to the number of the tetrahedra.  The entropy contributed by the macroscopic degeneracy has been measured experimentally , which is precisely predicted by Pauling who tries to explain the entropy of ice\cite{Pauling1935, Bramwell2001science}.

The monopole-like excitation is a single spin flip from one of the classical spin configuration of ground state.  The single spin flip breaks the rule of "2-in-2-out" in two adjacent tetrahedra creating a monopole-anti-monopole pair.  One tetrahedron becomes "1-in-3-out" spin configuration corresponding the monopole and the other becomes "3-in-1-out" corresponding the anti-monopole.  The monopole can hop away by flipping more spins.  If only the exchange interaction and the crystal field is taken into account, hopping monopole does not cost more energy, because there are only two tetrahedra breaking the rule of "2-in-2-out" in the hopping process.  In this case, the magnetic monopole is free after its nucleation.  If the dipolar interaction is further considered, numerical study shows that the interaction between monopoles becomes $1/r$, where $r$ is the bond length of the separation distance\cite{Castelnovo2007nature}.

The monopole picture seemingly explains the first order phase transition from the kagome ice state to the fully polarized spin state observed in Dy$_2$Ti$_2$O$_7$\cite{Sakakibara2003}.  As the magnetic field is applied in the [111] direction, the system undergoes a crossover to a partially polarized spin state, called the kagome ice state.  Along the [111] direction, the pyrochlore lattice can be viewed as a layered network of triangular lattice and kagome lattice alternating to each other.  The kagome ice state still obeys the rule of "2-in-2-out", but the spin orientations in the triangular layers are quenched.  Increasing the magnetic field in the kagome ice state generates monopole-anti-monopole pairs.  If it is dilute, it can be regarded as a gas phase of magnetic monopoles.  Further increment of concentration of monopole may lead to a liquid-gas phase transition to the fully-polarized state where all tetrahedra are in the spin orientation of "3-in-1-out"\cite{Sakakibara2003, aoki2004}.

The monopole picture to explain the magnetic properties of the spin ice material acquired some supports from the neutron scattering experiments\cite{Fennell2009nature, Morris2009science}.  An experiment of muon spin resonance, based on the magnetic Wien effect, indicates that the monopople-anti-monopole gas in Dy$_2$Ti$_2$O$_7$ behaves like an electrolyte\cite{bramwell2009nat}.  Recently, Dunsiger {\it et al.} showed negative result against the monopole signature in Dy$_2$Ti$_2$O$_7$\cite{dunsiger2011prl}.  Using the muon spin resonance, they measured the spin relaxation in the temperature region that monopole is believed to nucleate.  However, they obtained temperature independent spin relaxation, that is inconsistent with the nucleation of magnetic monopole in Dy$_2$Ti$_2$O$_7$\cite{dunsiger2011prl}.

The results by Dunsiger {\it et al.} actually shaken the foundation of the monopole picture in the spin ice material.  Although the monopole story is elegant and beautiful, it needs to be scrutinized, since the experiments supporting the monopole picture are in fact indirect evidence.  One of the theoretical supports for the monopole picture is the $1/r^3$ spin correlation in the spin ice state.  However, there is no direct experimental evidence supporting the existence of the state with the critical spin correlation at zero temperature.  On the contrary, there is experiment showing the spin freezing below $4K$ in Dy$_2$Ti$_2$O$_7$\cite{snyder2004}.  

In this paper, we report the calculation results of the spin structure factor and some thermodynamic quantities in the antiferromagnetic quantum Ising model in the pyrochlore lattice.  The Hamiltonian is given by the following
\begin{eqnarray}
H = J\sum_{\langle ij\rangle}\sigma^z_i\sigma^z_j -K\sum_i \sigma^x_i, \label{eq:hamiltonian}
\end{eqnarray}
where $\sigma^i$ are the Pauli spin matrices.  This Hamiltonian is relevant to the spin ice material because the effect of ferromagnetic exchange and the crystal field can be effectively represented by an antiferromagnetic Ising model.  Without the $K$ term, the Hamiltonian describes a classical spin disorder phase with $1/r^3$ spin correlation.  For finite $K$, the Hamiltonian describes a cooperative paramagnetic phase with exponentially-decayed spin correlation which adiabatically connects to the classical paramagnetic phase in the large $K$ limit\cite{chern2010}.  An important feature that Eq.~(\ref{eq:hamiltonian}) is different from the classical Hamiltonian is that the magnetic monopole is in the confined phase\cite{chern2010}.  Using a model with confined magnetic monopole and exponential spin correlation, we are able to reproduce the pinch point structure and its finite temperature properties that are observed in the neutron scattering experiment\cite{Fennell2009nature}.  The numerical results in the paper are calculated using the Monte Carlo simulation with the Trotter-Suzuki decomposition.  The cluster algorithm is applied in the imaginary time direction.  $10^{6}$ Monte Carlo steps with unbiased sampling and $10^5$ warm-up steps are used in the calculation.   This paper is organized as the following.  The finite temperature results of the pinch point is given in Section II.  The results of some thermodynamical quantities will be given in Section III.  In the Section IV, we conclude and the summarize the relevance of our calculation to the spin ice problem.

\section{Spin structure factor}\label{section:spin}

\begin{figure}[htb]
\includegraphics[width=0.5\textwidth]{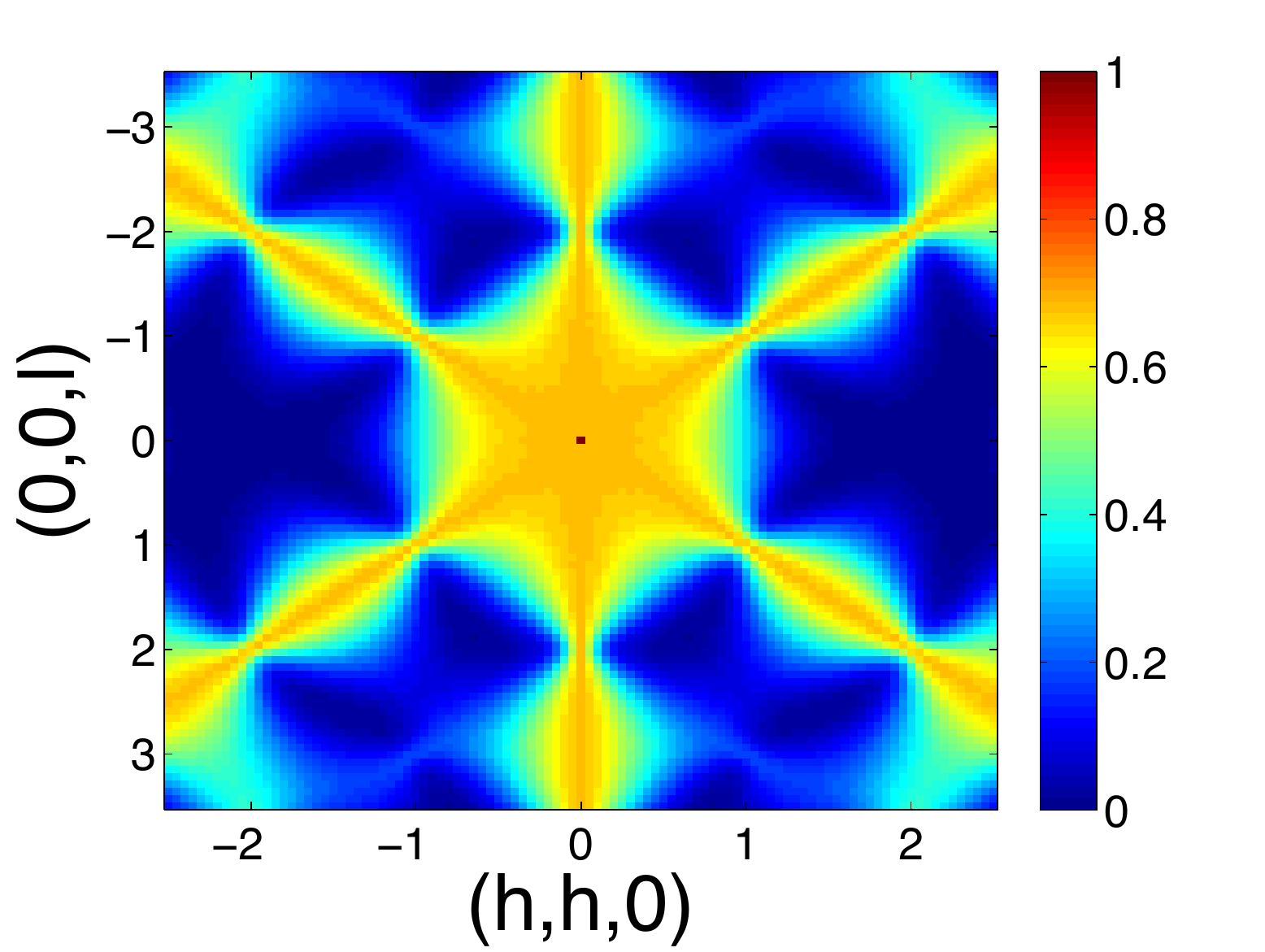}
\caption{(Color online) The spin structure factor in the spin-flip channel is calculated at $K=0.1$ and $T=0.6$ (in the unit of $J$).  The calculation is performed in the system consisting of $12\times12\times12$ cubic unit cells.  The pinch point structure at (0, 0, 2) is clearly seen.}\label{Fig:sf}
\end{figure}

The pinch point structure predicted theoretically\cite{henley2005} has been observed in the experiment of neutron scattering\cite{Fennell2009nature}.  It is a singularity in the spin structure factor in the spin flip channel.  At finite temperature, the singularity is broaden by thermal fluctuation and becomes Lorentzian distribution.  We compute the following spin structure factor in the pyrochlore lattice given by 
\begin{eqnarray}
I(\vec{Q})=\!\!\!\!\!\sum_{(i,a)(j,b)}\langle (\sigma^z)^a_i(\sigma^z)^b_j\rangle \cos(\vec{Q}\cdot \vec{R}^{ab}_{ij})(\hat{e}^a_{i,\perp}\cdot\hat{e}^{b}_{j,\perp}), \label{eq:structure}
\end{eqnarray}
where $<>$ is the thermal average, $a,b = 1,2,3,4$ denote the basis shown in Fig.~(\ref{Fig:pyrochlore}) and $i,j$ denote the fcc Bravais lattice and
\begin{eqnarray}
\vec{R}^{ab}_{ij}=\vec{r}^a_i-\vec{r}^b_j \ \ \ 
\hat{e}^a_{i,\perp}=\hat{e}^a_{i}-\frac{(\hat{e}^a_{i}\cdot \vec{Q})}{|\vec{Q}|^2}\vec{Q},
\end{eqnarray}
where $\hat{e}^a_i$ is unit vector of the local the easy axis.  The non spin-flip channel of the structure factor can be computed by
\begin{eqnarray}
\!\!\!\!\!I_{N\!S\!F}(\vec{Q})\!=\!\frac{2}{3}\!\!\!\sum_{(i,a)(j,b);a,b> 2}\!\!\!\!\!\!\!\!\!\!\!\!\!\langle(\sigma^z)^a_i(\sigma^z)^b_j\rangle\! \cos(\vec{Q}\!\cdot\! \vec{R}^{ab}_{ij})(2\delta_{ab}\!-\!1).
\end{eqnarray}
The spin-flip channel $I_{SF}(\vec{Q})$ is obtained by subtracting $I_{NSF}(\vec{Q})$ from $I(\vec{Q})$, $I_{SF}(\vec{Q})=I(\vec{Q})-I_{NSF}(\vec{Q})$.  A result of the spin structure factor in the spin-flip channel at $K=0.1$ and $T=0.6$ in the lattice of $12\times12\times12$ unit cells is shown in Fig.~(\ref{Fig:sf}).  Since there are 16 spins in a unit cell, the calculation in Fig.~(\ref{Fig:sf}) is performed in the system of 27648 sites.

\begin{figure}[htb]
\includegraphics[width=0.5\textwidth]{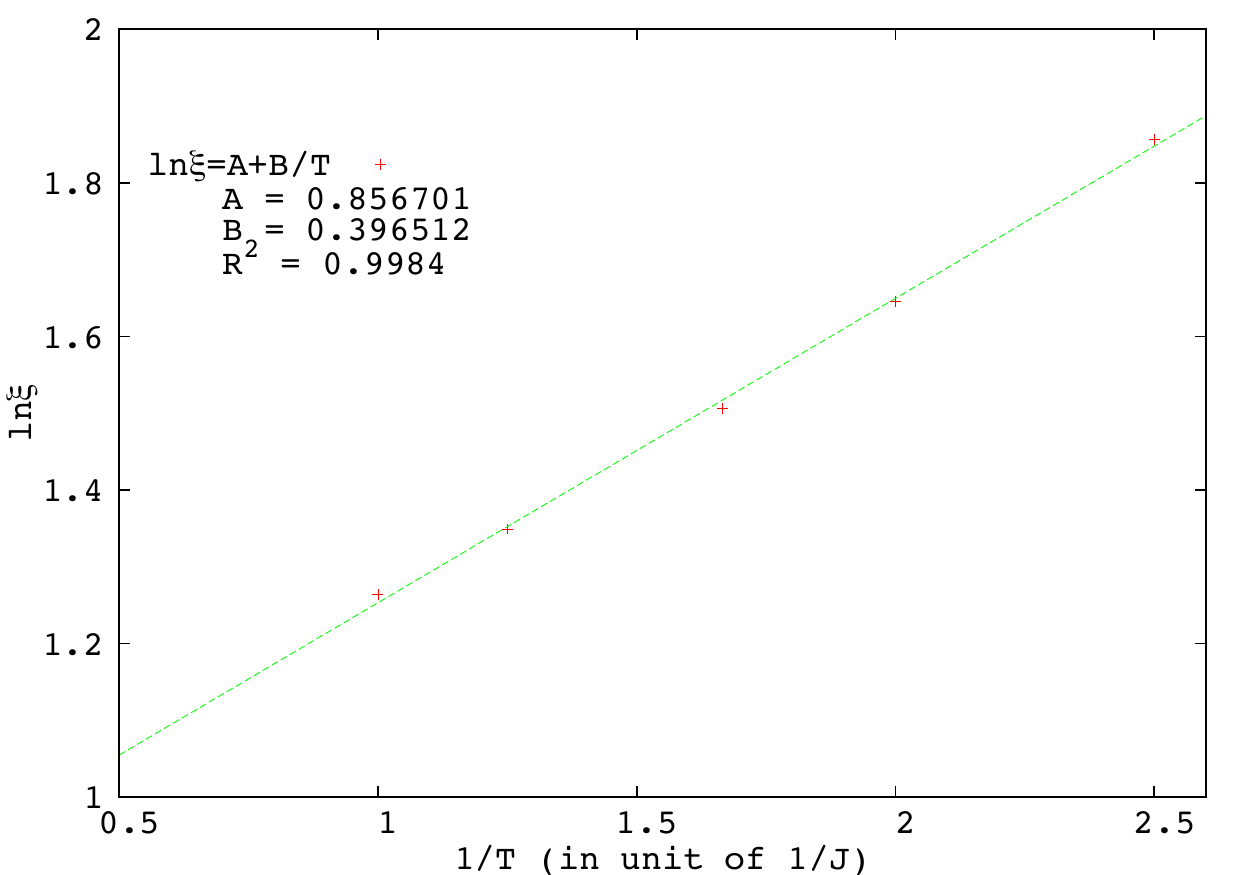}
\caption{(Color online) The temperature dependence of the width of the Lorentzian distribution at the pinch point.  The calculation is performed for 5 different values of temperature down to $T=0.4$.    $\xi(T)\sim$ exp$(B/T)$, where $B\approx0.4$, with the R$^2$ value 0.9984.  }\label{Fig:xi}
\end{figure}

Our result of the spin structure factor in the spin-flip channel looks very similar to the experimental results by Fennell {\it et al.}\cite{Fennell2009nature}.  In Fig.~(\ref{Fig:sf}), a pinch point at $(0,0,2)$ is clearly observed.  Due to the thermal fluctuation, the pinch point is broaden described by a Lorentzian distribution.  We analyze the temperature dependence of the width of Lorentzian distribution $\xi^{-1}$, which can be extracted from the computational data.  The results are given in Fig.~(\ref{Fig:xi}).  The finite size effect is carefully handled.  We compute for the systems of many sizes until it is large enough so that the value of $\xi$ is saturated for each temperature.  The maximum number of spin computed is 27648 in the $12\times12\times12$ unit cells with period boundary condition.  Our results of the temperature dependence of $\xi$ is consistent with the experimental results\cite{Fennell2009nature}.  $\xi(T)\sim$ exp$(B/T)$, where $B\approx0.4$.

Our results indicate that the critical spin correlation is not a necessary condition for the existence of the pinch point structure in the neutron scattering data.  Our model that exhibits exponentially-decayed spin correlation\cite{chern2010} provides the same structure with the accurate temperature dependence.  Although $\xi$ has a dimension of length, it is not the typical  spin correlation length defined in the textbook of magnetism.  As a measure of the thermal fluctuation, the thermal broadening of the pinch point is similar to the one of the Bragg peak in the neutron scattering.  Therefore, the divergence of the $\xi$ does not imply the divergence of the spin correlation length, {\it i.e.} a critical phase, at zero temperature.

\section{Thermodynamics}\label{section:thermodynamics}
\subsection{Magnetization}
\begin{figure}[htb]
\includegraphics[width=0.5\textwidth]{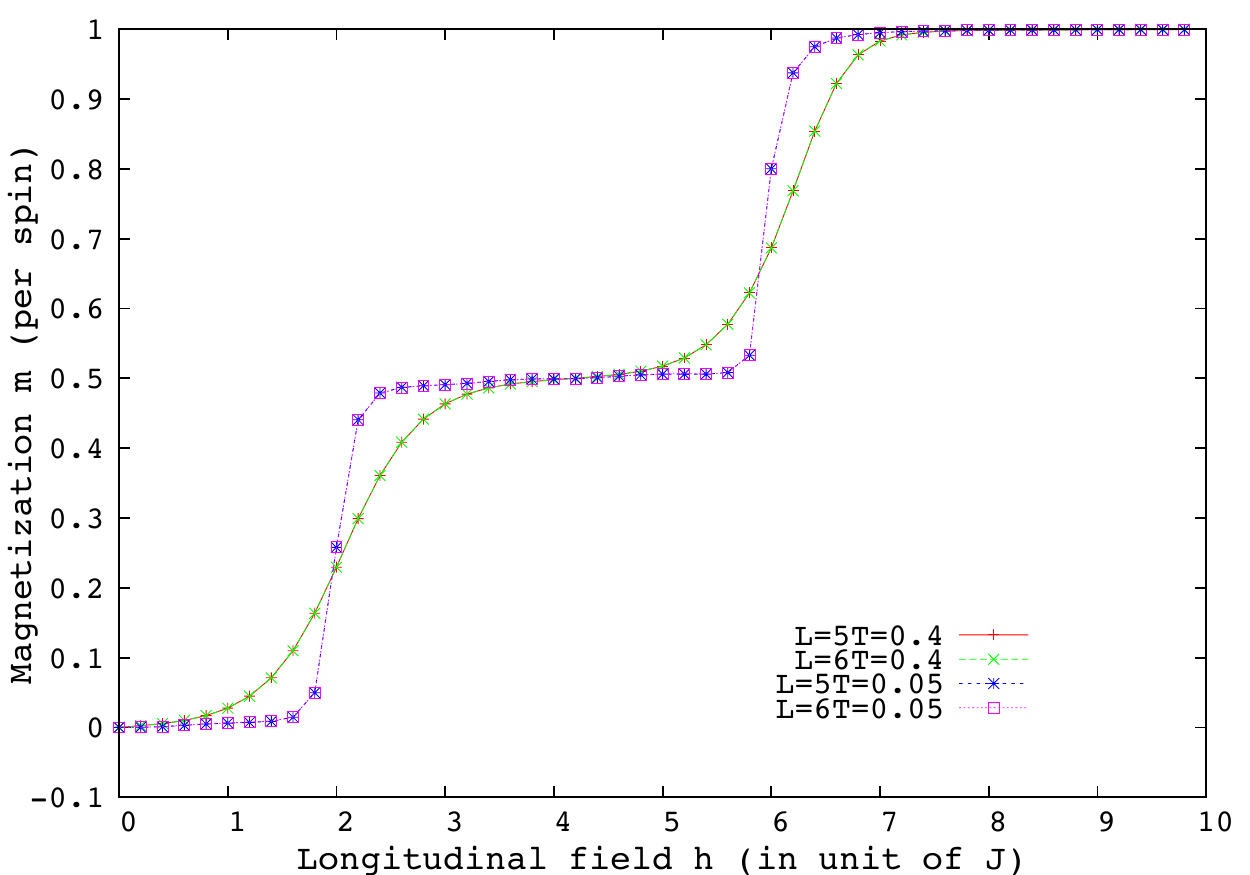}
\caption{(Color online) The field dependence of the magnetization at $K=0.1$.  The phase diagram is characterized by three phases: spin disorder phase at zero field, kagome ice phase at intermediate field, and the fully-polarized state in the high field.  The plateau structure is also observed in experiments.  Two steps of transition between those states are crossover similar to the classical nearest-neighbor model for the spin ice material.}\label{Fig:mag}
\end{figure}

In Fig.~(\ref{Fig:mag}), we plot the field dependence of the magnetization for $K=0.1$ at two different temperature $T = 0.4$ and $T = 0.05$.  The field of strength $h$ is applied in the longitudinal direction.  The Hamiltonian is described by
\begin{eqnarray}
H = J\sum_{\langle ij\rangle}\sigma^z_i\sigma^z_j -K\sum_i \sigma^x_i- h \sum_i \sigma^z_i \label{eq:hamimag}
\end{eqnarray}

For each temperature, we show the results of two system sizes of $L\times L\times L$ unit cells for $L=5$ and 6.   The results of $L=5$ and $L=6$ are almost identical at both temperature.  Namely, thermodynamic limit is reached.  At zero field, the magnetization is zero and the system is spin disorder consistent with disorder by disorder\cite{chern2010}.  As the field is applied, the system undergoes a crossover to state of magnetization $m=0.5$ per site.  Because the Ising variable is 1 or $-1$ at each site, the state of $m=0.5$ is the one that there are three sites with $\sigma=1$ and one site with $\sigma=-1$ in every tetrahedron.  Restoring the spin orientations by identifying the local [111] axis, it can be shown that the state of $m=0.5$ corresponds to the "2-in-2-out" kagome ice state.  The crossover from a spin disorder phase to the kagome ice state is already observed in the experiment\cite{Sakakibara2003}.  

The crossover can be understood by the nearest-neighbor model of spin ice.  Both the $m=0$ and the $m=0.5$ are the phases with spin gap.  The field in the [111] direction does not close the gap, and therefore the transition is a crossover.  Since the kagome ice state is a gapped state, it is robust against the magnetic field.  The field dependence of the magnetization exhibits a clear plateau structure.  The width of the plateau depends on the temperature.  The lower the temperature is, the wider the plateau becomes.

As field is further applied, the system undergoes another crossover from the kagome state to the fully-polarized state at both $T=0.4$ and $T=0.05$.  It is a state of the Ising variable $\sigma=1$ at all spin sites.  Restoring the spin orientation, it corresponds to "3-in-1-out" spin polarized state.  Our result is consistent with the nearest-neighbor model but cannot explain the first order phase transition observed in the experiment\cite{Sakakibara2003}.  In the experiment\cite{Sakakibara2003,aoki2004}, the nature is the transition is different at $T=0.4$ and $T=0.05.$  It is a crossover at $T=0.4$, and it is a first order phase transition at $T=0.05$.  The first order phase transition is believed to the attributed by  the dipolar interaction\cite{Castelnovo2007nature} as claimed in Ref.[5].

\subsection{Magnetic susceptibility}

\begin{figure}[htb]
\includegraphics[width=0.5\textwidth]{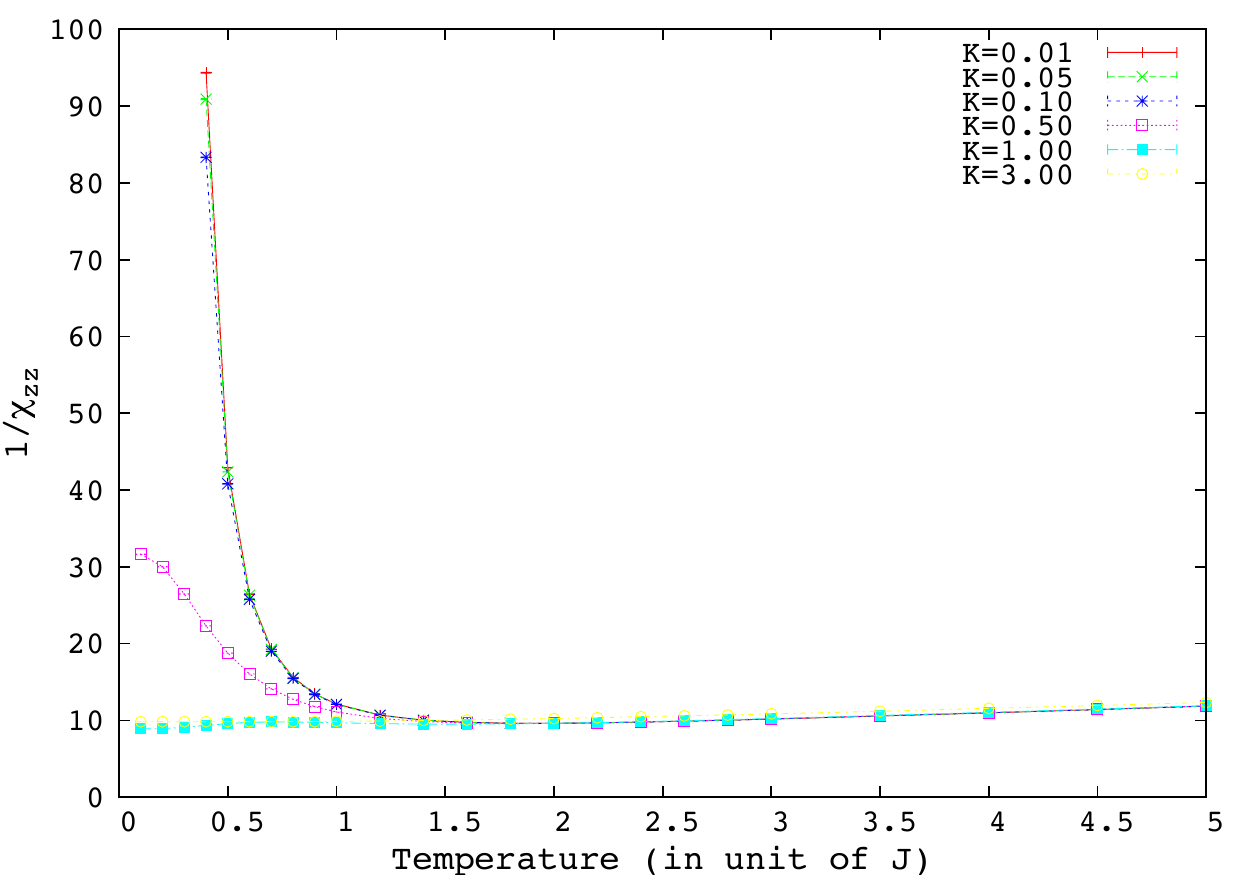}
\caption{(Color online) The temperature dependence of the inverse of the longitudinal magnetic susceptibility for $K=0.01$, 0.05, 0.1, 1, and 3 in the system of $10\times10\times10$ unit cells.}\label{Fig:inverse}
\end{figure}

In Fig.~(\ref{Fig:inverse}), we compute for the temperature dependence of the inverse of the longitudinal susceptibility.  The calculation is taken in the system of $10\times10\times10$ unit cells containing 16000 spins for 6 different values of $K$.  In the high temperature region, all systems have the same susceptibility obeying the Curie-Weiss law.  Since the magnetic susceptibility goes to zero at zero temperature  in the $K=0$ case, the inverse of the susceptibility deviates from the Curie-Weiss law and diverges in the low temperature.  For large $K$, the magnetic susceptibility obeys the Curie-Weiss law perfectly.  The susceptibility at zero temperature becomes finite for finite $K$.  However, due to the numerical precision, this statement is firmly confirmed only for $K\ge0.05$.  For $K= 0.01$, the susceptibility in the low temperature is too small beyond the precision of the machine.  Similar to the antiferromagnetic quantum Ising model in the kagome lattice\cite{chern2008}, the smooth transformation of the magnetic susceptibility in the variation of $K$ is consistent with the analysis in Ref.[11].  Namely, the system undergoes a crossover from the cooperative paramagnetic state to the classical paramagnetic state in the large $K$ limit.  

\subsection{Specific heat and entropy}

\begin{figure}[htb]
\includegraphics[width=0.5\textwidth]{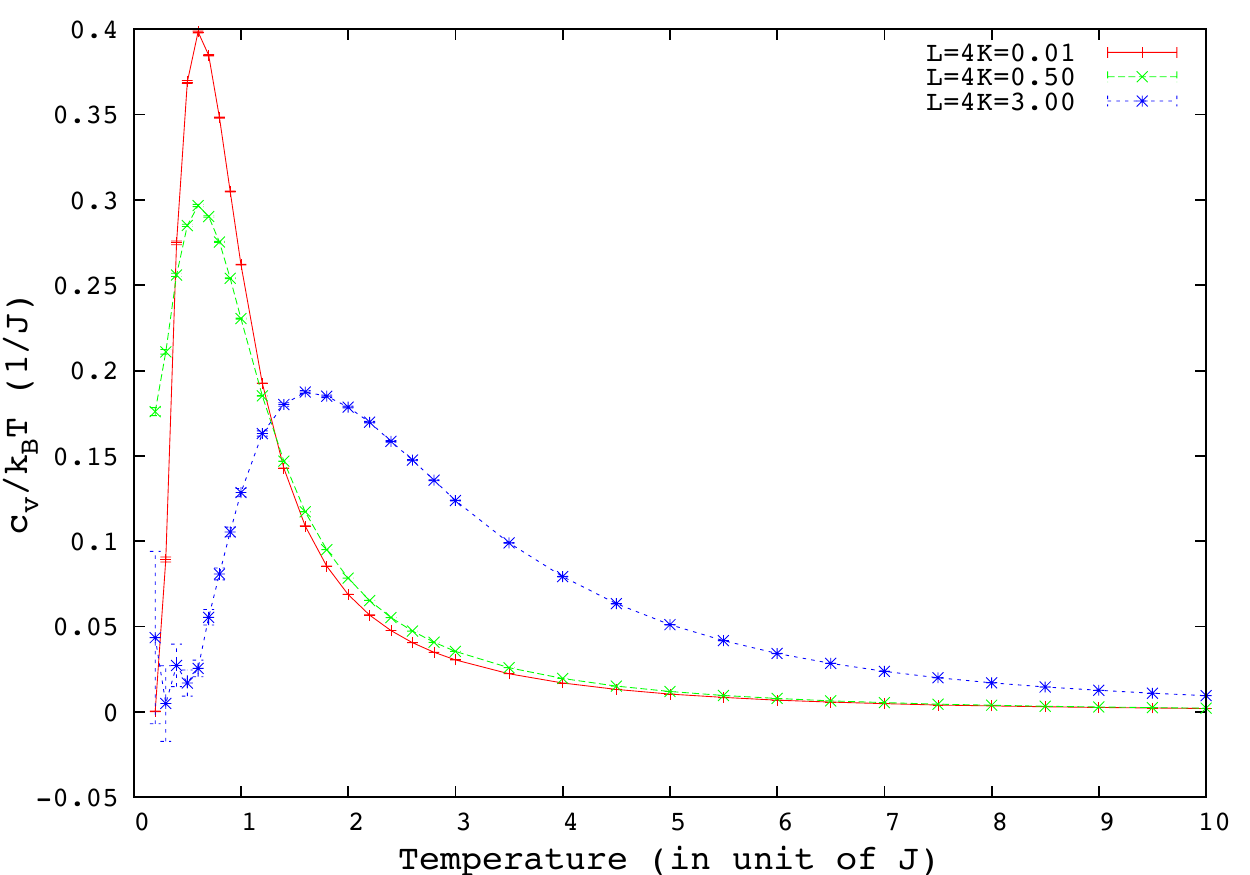}
\includegraphics[width=0.5\textwidth]{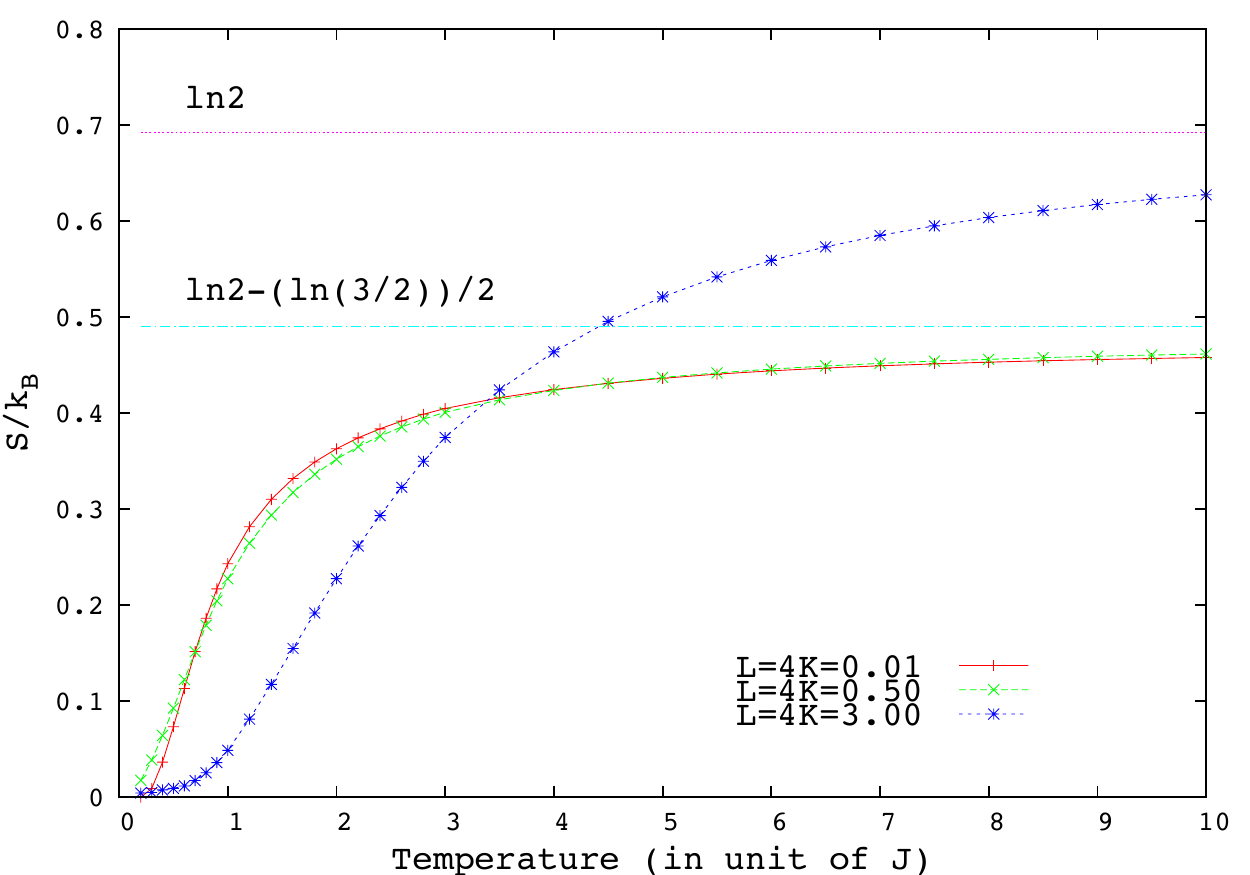}
\caption{(Color online) (a) The temperature dependence of Cv/$T$ for $K=0.01$, 0.05, and 3.  For the purpose of integration, the data of many values of temperature is needed.  Therefore, the calculation is performed in the system of $4\times4\times 4$ unit cells for efficiency.  (b) Temperature dependence of the entropy.}\label{Fig:cvdt}
\end{figure}

In Fig.~(\ref{Fig:cvdt}a), we show the results of temperature dependence of Cv/$T$ for $K=0.01$, 0.05, and 3.  The calculation is performed for the temperature down to $T=0.1$ in the systems of $4\times4\times4$ unit cells.  The finite size effect is carefully considered.  Although the results are not shown here, the system of $4\times4\times4$ unit cells is large enough representing the thermodynamic limit.  The entropy is also computed.  In Fig.~(\ref{Fig:cvdt}b), we show the results of the temperature dependence of the entropy.  For $K=0.01$ and 0.05, the entropy in high temperature saturate at $\ln2-\frac{1}{2}\ln\frac{3}{2}$, where $\frac{1}{2}\ln\frac{3}{2}$ is the residual entropy of the spin ice state.

For finite $K$, the macroscopic ground state degeneracy is lifted due to the quantum fluctuation.  The value of $K$ introduces an energy scale of lifting.  As the lowest temperature to be $T=0.1$ is considered, the quantum Ising model gives the same results as the classical model\cite{Bramwell2001science} for $K=0.01$ and 0.05.  We note that the error of entropy mainly comes from the specific heat in the low temperature.  Not only the specific heat has larger error in the low temperature, but also the variation is sharper and requires finer grids of temperature in the integration of Cv/$T$.

For $K=3$, the system becomes classical paramagnetic phase.  The entropy starts from zero and saturates at $\ln2$ as temperature increases.  Due to the limitation of the algorithm, the error of specific heat is significant in low temperature, because the specific heat decreases to zero exponentially as temperature approaches to zero.  Numerically, it causes significant error.  On the other hand, the specific heat at $K=3$ decreases to zero slowly in the high temperature, leading to some error in the calculation of entropy.

\section{Discussion and conclusion}\label{section:discussion}

In this section, we discuss the relevancy of our calculation to the physics of spin ice material.  First of all, we stress that we are not aiming at explaining all the experimental discovery in the spin ice system.  The dipolar interaction which plays a significant role in the real systems is neglected here.  The purpose of this paper is to provide a model of monopole confinement to reproduce experimental results that are pronounced to be the evidence of deconfinement of magnetic monopole.

The existence of the pinch point structure in the spin structure factor in the spin flip channel is thought to support the existence of the critical spin ice state.  However, our results show that the thermal broadening at the pinch points is not different from the thermal broadening at the Bragg peak measured at finite temperature.  The vanishing of the width of the Lorentzian distribution does not imply the divergence of the spin correlation length.  Furthermore, in the experiment of Chang {\it et al.}\cite{chang2010}, the pinch point structure is shown to be independent of the doping of non-magnetic ions.  In the Ho$_{2-x}$Y$_x$Ti$_2$O$_7$, the magnitude of the pinch point is normalized showing almost identical structure for $x=0$, 0.3, and 1. It is well known that the doping effect usually changes the spin correlation dramatically.  One of the best examples is the high temperature superconductor.  There are also numerous transition metal oxides exhibiting diverse phase diagram due to doping, especially when the system is in the criticality.  There are two possible scenarios here: 1) If Ho$_{2}$Ti$_2$O$_7$ is in the critical spin ice state, the spin correlation should change dramatically due to the doping.  At the 50\% doping, where the system is not critical, the pinch point structure is still robust\cite{chang2010}.  2) Ho$_{2}$Ti$_2$O$_7$ is not in the criticality.  Doping with non-magnetic ion, the system is not critical, too.  In both cases, the existence of the pinch point structure does not imply the criticality at zero temperature.

Our result of the specific heat is also consistent with Bramwell {\it et al.}\cite{Bramwell2001science}.  The saturation of the entropy reaches $\ln 2-\frac{1}{2}\ln\frac{3}{2}$ at high temperature.  The missing entropy is contributed by the macroscopic ground state degeneracy but lifted by the quantum fluctuation.

In the field dependent magnetization, the longitudinal magnetic field is applied to the local [111] direction.  Since there are four different local axes in the fundamental tetrahedra, there are four local [111] directions different site by site in one tetrahedron.  Although our result of field dependent does not have experimental realization, it serves as an important result comparing with the classical nearest-neighbor model of spin ice.

The cooperative paramagnetic state at finite but small $K$ can be adiabatically connected to the classical paramagnetic state in the large $K$ limit\cite{chern2010}.  Our calculation of the temperature dependence of the magnetic susceptibility for various $K$ is consistent with that result.

Finally, our model at small $K$, exhibiting the cooperative paramagnetic phase with confined magnetic monopole, has the same spin structure factor observed in the experiments, giving another counter example for the claimed existence of magnetic monopole.  As suggested by Dunsiger {\it et al.}, we may need another term in the Hamiltonian to describe real spin ice material.  Our results suggest that the quantum fluctuation should be seriously considered.

We heartily appreciate the stimulated discussion with Ying-Jer Kao, Naoto Nagaosa and Dung-Hai Lee.  This work is supported by National Science Council of Taiwan under NSC 97-2112-M-002-027-MY3 and NSC 100-2112-M-002-015-MY3.

\maketitle



\begin{thebibliography}{10}%
\makeatletter
\providecommand \@ifxundefined [1]{%
 \ifx #1\undefined \expandafter \@firstoftwo
 \else \expandafter \@secondoftwo
\fi
}%
\providecommand \@ifnum [1]{%
 \ifnum #1\expandafter \@firstoftwo
 \else \expandafter \@secondoftwo
\fi
}%
\providecommand \enquote [1]{``#1''}%
\providecommand \bibnamefont  [1]{#1}%
\providecommand \bibfnamefont [1]{#1}%
\providecommand \citenamefont [1]{#1}%
\providecommand\href[0]{\@sanitize\@href}%
\providecommand\@href[1]{\endgroup\@@startlink{#1}\endgroup\@@href}%
\providecommand\@@href[1]{#1\@@endlink}%
\providecommand \@sanitize [0]{\begingroup\catcode`\&12\catcode`\#12\relax}%
\@ifxundefined \pdfoutput {\@firstoftwo}{%
 \@ifnum{\z@=\pdfoutput}{\@firstoftwo}{\@secondoftwo}%
}{%
 \providecommand\@@startlink[1]{\leavevmode}%
 \providecommand\@@endlink[0]{}%
}{%
 \providecommand\@@startlink[1]{%
  \leavevmode
  \pdfstartlink
   attr{/Border[0 0 1 ]/H/I/C[0 1 1]}%
   user{/Subtype/Link/A<</Type/Action/S/URI/URI(#1)>>}%
  \relax
 }%
 \providecommand\@@endlink[0]{\pdfendlink}%
}%
\providecommand \url  [0]{\begingroup\@sanitize \@url }%
\providecommand \@url [1]{\endgroup\@href {#1}{\urlprefix}}%
\providecommand \urlprefix [0]{URL }%
\providecommand \Eprint[0]{\href }%
\@ifxundefined \urlstyle {%
  \providecommand \doi [1]{doi:\discretionary{}{}{}#1}%
}{%
  \providecommand \doi [0]{doi:\discretionary{}{}{}\begingroup
  \urlstyle{rm}\Url }%
}%
\providecommand \doibase [0]{http://dx.doi.org/}%
\providecommand \Doi[1]{\href{\doibase#1}}%
\providecommand \bibAnnote [3]{%
  \BibitemShut{#1}%
  \begin{quotation}\noindent
    \textsc{Key:}\ #2\\\textsc{Annotation:}\ #3%
  \end{quotation}%
}%
\providecommand \bibAnnoteFile [2]{%
  \IfFileExists{#2}{\bibAnnote {#1} {#2} {\input{#2}}}{}%
}%
\providecommand \typeout [0]{\immediate \write \m@ne }%
\providecommand \selectlanguage [0]{\@gobble}%
\providecommand \bibinfo [0]{\@secondoftwo}%
\providecommand \bibfield [0]{\@secondoftwo}%
\providecommand \translation [1]{[#1]}%
\providecommand \BibitemOpen[0]{}%
\providecommand \bibitemStop [0]{}%
\providecommand \bibitemNoStop [0]{.\EOS\space}%
\providecommand \EOS [0]{\spacefactor3000\relax}%
\providecommand \BibitemShut [1]{\csname bibitem#1\endcsname}%
\bibitem{snyder2004}%
  \BibitemOpen
  \bibfield{author}{%
  \bibinfo {author} {\bibfnamefont{J.}~\bibnamefont{Snyder}}, \bibinfo {author}
  {\bibfnamefont{B.~G.}\ \bibnamefont{Ueland}}, \bibinfo {author}
  {\bibfnamefont{J.~S.}\ \bibnamefont{Slusky}}, \bibinfo {author}
  {\bibfnamefont{H.}~\bibnamefont{Karunadasa}}, \bibinfo {author}
  {\bibfnamefont{R.~J.}\ \bibnamefont{Cava}},\ and\ \bibinfo {author}
  {\bibfnamefont{P.}~\bibnamefont{Schiffer}},\ }%
  \bibfield{journal}{%
  \bibinfo {journal} {Phys. Rev. B}\ }%
  \textbf{\bibinfo {volume} {69}},\ \bibinfo {pages} {064414} (\bibinfo {year}
  {2004})%
  \bibAnnoteFile{NoStop}{snyder2004}%
\bibitem{Pauling1935}%
  \BibitemOpen
  \bibfield{author}{%
  \bibinfo {author} {\bibfnamefont{L.}~\bibnamefont{Pauling}},\ }%
  \bibfield{journal}{%
  \bibinfo {journal} {Journal of the American Chemical Socienty}\ }%
  \textbf{\bibinfo {volume} {57}},\ \bibinfo {pages} {2680} (\bibinfo {year}
  {1935})%
  \bibAnnoteFile{NoStop}{Pauling1935}%
\bibitem{Bramwell2001science}%
  \BibitemOpen
  \bibfield{author}{%
  \bibinfo {author} {\bibfnamefont{S.}~\bibnamefont{Bramwell}}\ and\ \bibinfo
  {author} {\bibfnamefont{M.~J.~P.}\ \bibnamefont{Gingras}},\ }%
  \bibfield{journal}{%
  \bibinfo {journal} {Science}\ }%
  \textbf{\bibinfo {volume} {294}},\ \bibinfo {pages} {1495} (\bibinfo {year}
  {2001})%
  \bibAnnoteFile{NoStop}{Bramwell2001science}%
\bibitem{Castelnovo2007nature}%
  \BibitemOpen
  \bibfield{author}{%
  \bibinfo {author} {\bibfnamefont{C.}~\bibnamefont{Castelnovo}}, \bibinfo
  {author} {\bibfnamefont{R.}~\bibnamefont{Moessner}},\ and\ \bibinfo {author}
  {\bibfnamefont{S.}~\bibnamefont{Sondhi}},\ }%
  \bibfield{journal}{%
  \bibinfo {journal} {Nature}\ }%
  \textbf{\bibinfo {volume} {451}},\ \bibinfo {pages} {42} (\bibinfo {year}
  {2007})%
  \bibAnnoteFile{NoStop}{Castelnovo2007nature}%
\bibitem{Sakakibara2003}%
  \BibitemOpen
  \bibfield{author}{%
  \bibinfo {author} {\bibfnamefont{T.}~\bibnamefont{Sakakibara}}, \bibinfo
  {author} {\bibfnamefont{T.}~\bibnamefont{Tayama}}, \bibinfo {author}
  {\bibfnamefont{Z.}~\bibnamefont{Hiroi}}, \bibinfo {author}
  {\bibfnamefont{K.}~\bibnamefont{Matsuhira}},\ and\ \bibinfo {author}
  {\bibfnamefont{S.}~\bibnamefont{Takagi}},\ }%
  \bibfield{journal}{%
  \bibinfo {journal} {Phys. Rev. Lett.}\ }%
  \textbf{\bibinfo {volume} {90}},\ \bibinfo {pages} {207205} (\bibinfo {year}
  {2003})%
  \bibAnnoteFile{NoStop}{Sakakibara2003}%
\bibitem{aoki2004}%
  \BibitemOpen
  \bibfield{author}{%
  \bibinfo {author} {\bibfnamefont{H.}~\bibnamefont{Aoki}}, \bibinfo {author}
  {\bibfnamefont{T.}~\bibnamefont{Sakakibara}}, \bibinfo {author}
  {\bibfnamefont{K.}~\bibnamefont{Matsuhira}},\ and\ \bibinfo {author}
  {\bibfnamefont{Z.}~\bibnamefont{Hiroi}},\ }%
  \bibfield{journal}{%
  \bibinfo {journal} {J. Phys. Soc. Japan}\ }%
  \textbf{\bibinfo {volume} {73}},\ \bibinfo {pages} {2851} (\bibinfo {year}
  {2004})%
  \bibAnnoteFile{NoStop}{aoki2004}%
\bibitem{Fennell2009nature}%
  \BibitemOpen
  \bibfield{author}{%
  \bibinfo {author} {\bibfnamefont{T.}~\bibnamefont{Fennell}}, \bibinfo
  {author} {\bibfnamefont{P.}~\bibnamefont{Deen}}, \bibinfo {author}
  {\bibfnamefont{A.}~\bibnamefont{Wildes}}, \bibinfo {author}
  {\bibfnamefont{K.}~\bibnamefont{Schmalzl}}, \bibinfo {author}
  {\bibfnamefont{D.}~\bibnamefont{Prabhakaran}}, \bibinfo {author}
  {\bibfnamefont{A.}~\bibnamefont{Boothroyd}}, \bibinfo {author}
  {\bibfnamefont{R.}~\bibnamefont{Aldus}}, \bibinfo {author}
  {\bibfnamefont{D.}~\bibnamefont{McMorrow}},\ and\ \bibinfo {author}
  {\bibfnamefont{S.}~\bibnamefont{Bramwell}},\ }%
  \bibfield{journal}{%
  \bibinfo {journal} {Science}\ }%
  \textbf{\bibinfo {volume} {326}},\ \bibinfo {pages} {415} (\bibinfo {year}
  {2009})%
  \bibAnnoteFile{NoStop}{Fennell2009nature}%
\bibitem{Morris2009science}%
  \BibitemOpen
  \bibfield{author}{%
  \bibinfo {author} {\bibfnamefont{D.}~\bibnamefont{Morris}}, \bibinfo {author}
  {\bibfnamefont{D.}~\bibnamefont{Tennant}}, \bibinfo {author}
  {\bibfnamefont{S.}~\bibnamefont{Grigera}}, \bibinfo {author}
  {\bibfnamefont{B.}~\bibnamefont{Klemke}}, \bibinfo {author}
  {\bibfnamefont{C.}~\bibnamefont{Castelnovo}}, \bibinfo {author}
  {\bibfnamefont{R.}~\bibnamefont{Moessner}}, \bibinfo {author}
  {\bibfnamefont{C.}~\bibnamefont{Czternasty}}, \bibinfo {author}
  {\bibfnamefont{M.}~\bibnamefont{Meissner}}, \bibinfo {author}
  {\bibfnamefont{K.}~\bibnamefont{Rule}}, \bibinfo {author}
  {\bibfnamefont{J.-U.}\ \bibnamefont{Hoffmann}}, \bibinfo {author}
  {\bibfnamefont{K.}~\bibnamefont{Kiefer}}, \bibinfo {author}
  {\bibfnamefont{S.}~\bibnamefont{Gerischer}}, \bibinfo {author}
  {\bibfnamefont{D.}~\bibnamefont{Slobinsky}},\ and\ \bibinfo {author}
  {\bibfnamefont{R.~S.}\ \bibnamefont{Perry}},\ }%
  \bibfield{journal}{%
  \bibinfo {journal} {Science}\ }%
  \textbf{\bibinfo {volume} {326}},\ \bibinfo {pages} {411} (\bibinfo {year}
  {2009})%
  \bibAnnoteFile{NoStop}{Morris2009science}%
\bibitem{bramwell2009nat}%
  \BibitemOpen
  \bibfield{author}{%
  \bibinfo {author} {\bibfnamefont{S.}~\bibnamefont{Bramwell}}, \bibinfo
  {author} {\bibfnamefont{S.}~\bibnamefont{Giblin}}, \bibinfo {author}
  {\bibfnamefont{S.}~\bibnamefont{Calder}}, \bibinfo {author}
  {\bibfnamefont{R.}~\bibnamefont{Aldus}}, \bibinfo {author}
  {\bibfnamefont{D.}~\bibnamefont{Prabhakaran}},\ and\ \bibinfo {author}
  {\bibfnamefont{T.}~\bibnamefont{Fennell}},\ }%
  \bibfield{journal}{%
  \bibinfo {journal} {Nature}\ }%
  \textbf{\bibinfo {volume} {461}},\ \bibinfo {pages} {956} (\bibinfo {year}
  {2009})%
  \bibAnnoteFile{NoStop}{bramwell2009nat}%
\bibitem{dunsiger2011prl}%
  \BibitemOpen
  \bibfield{author}{%
  \bibinfo {author} {\bibfnamefont{S.~R.}\ \bibnamefont{Dunsiger}}, \bibinfo
  {author} {\bibfnamefont{A.~A.}\ \bibnamefont{Aczel}}, \bibinfo {author}
  {\bibfnamefont{C.}~\bibnamefont{Arguello}}, \bibinfo {author}
  {\bibfnamefont{H.}~\bibnamefont{Dabkowska}}, \bibinfo {author}
  {\bibfnamefont{A.}~\bibnamefont{Dabkowski}}, \bibinfo {author}
  {\bibfnamefont{M.-H.}\ \bibnamefont{Du}}, \bibinfo {author}
  {\bibfnamefont{T.}~\bibnamefont{Goko}}, \bibinfo {author}
  {\bibfnamefont{B.}~\bibnamefont{Javanparast}}, \bibinfo {author}
  {\bibfnamefont{T.}~\bibnamefont{Lin}}, \bibinfo {author}
  {\bibfnamefont{F.~L.}\ \bibnamefont{Ning}}, \bibinfo {author}
  {\bibfnamefont{H.~M.~L.}\ \bibnamefont{Noad}}, \bibinfo {author}
  {\bibfnamefont{D.~J.}\ \bibnamefont{Singh}}, \bibinfo {author}
  {\bibfnamefont{T.~J.}\ \bibnamefont{Williams}}, \bibinfo {author}
  {\bibfnamefont{Y.~J.}\ \bibnamefont{Uemura}}, \bibinfo {author}
  {\bibfnamefont{M.~J.~P.}\ \bibnamefont{Gingras}},\ and\ \bibinfo {author}
  {\bibfnamefont{G.~M.}\ \bibnamefont{Luke}},\ }%
  \bibinfo {journal} {arXiv:1110.0877, accepted for publication in Phys. Rev.
  Lett.}%
  \bibAnnoteFile{Stop}{dunsiger2011prl}%
\bibitem{chern2010}%
  \BibitemOpen
\bibfield{journal}{%
    }%
  \bibfield{author}{%
  \bibinfo {author} {\bibfnamefont{C.-H.}\ \bibnamefont{Chern}}\ and\ \bibinfo
  {author} {\bibfnamefont{C.-N.}\ \bibnamefont{Liao}},\ }%
  \bibinfo {journal} {arXiv:1003.4204}%
  \bibAnnoteFile{NoStop}{chern2010}%
\bibitem{henley2005}%
  \BibitemOpen
\bibfield{journal}{%
    }%
  \bibfield{author}{%
  \bibinfo {author} {\bibfnamefont{C.}~\bibnamefont{Henley}},\ }%
  \bibfield{journal}{%
  \bibinfo {journal} {Phys. Rev. B}\ }%
  \textbf{\bibinfo {volume} {71}},\ \bibinfo {pages} {014424} (\bibinfo {year}
  {2005})%
  \bibAnnoteFile{NoStop}{henley2005}%
\bibitem{chang2010}%
  \BibitemOpen
  \bibfield{author}{%
  \bibinfo {author} {\bibfnamefont{L.~J.}\ \bibnamefont{Chang}}, \bibinfo
  {author} {\bibfnamefont{Y.}~\bibnamefont{Su}}, \bibinfo {author}
  {\bibfnamefont{Y.~J.}\ \bibnamefont{Kao}}, \bibinfo {author}
  {\bibfnamefont{Y.~Z.}\ \bibnamefont{Chou}}, \bibinfo {author}
  {\bibfnamefont{R.}~\bibnamefont{Mittal}}, \bibinfo {author}
  {\bibfnamefont{H.}~\bibnamefont{Schneider}}, \bibinfo {author}
  {\bibfnamefont{T.}~\bibnamefont{Brueckel}}, \bibinfo {author}
  {\bibfnamefont{G.}~\bibnamefont{Balakrishnan}},\ and\ \bibinfo {author}
  {\bibfnamefont{M.~R.}\ \bibnamefont{Lees}},\ }%
  \bibfield{journal}{%
  \bibinfo {journal} {Phys. Rev. B}\ }%
  \textbf{\bibinfo {volume} {82}},\ \bibinfo {pages} {172403} (\bibinfo {year}
  {2010})%
  \bibAnnoteFile{NoStop}{chang2010}%
\bibitem{chern2008}%
  \BibitemOpen
  \bibfield{author}{%
  \bibinfo {author} {\bibfnamefont{C.-H.}\ \bibnamefont{Chern}}\ and\ \bibinfo
  {author} {\bibfnamefont{M.}~\bibnamefont{Tsukamoto}},\ }%
  \bibfield{journal}{%
  \bibinfo {journal} {Phys. Rev. B}\ }%
  \textbf{\bibinfo {volume} {77}},\ \bibinfo {pages} {172404} (\bibinfo {year}
  {2008})%
  \bibAnnoteFile{NoStop}{chern2008}%
\end{thebibliography}
%

\end{document}